\pdfoutput=1

\documentclass[11pt]{article}

\usepackage[preprint]{acl}

\usepackage{times}
\usepackage{latexsym}   
\usepackage[T1]{fontenc}

\usepackage[utf8]{inputenc}

\usepackage{enumitem} 
\usepackage{microtype}

\usepackage{inconsolata}

\usepackage{graphicx}

\usepackage{booktabs}
\usepackage{subcaption}
\usepackage{caption}
\usepackage{listings}
\usepackage{hyperref}

\usepackage{amsmath,amsfonts,bm}
\usepackage{soul}
\usepackage{afterpage}
\newcommand{\sysName}{{{ChatWise}}} 
\newcommand{\J}[1]{\textcolor{black}{#1}} 
\newcommand{\Zhengbang}[1]{\textcolor{black}{#1}} 

\def\ie{{\textit{i.e.}}}
\def\eg{{\textit{e.g.}}}

\title{ChatWise: A Strategy-Guided Chatbot for Enhancing Cognitive Support in Older Adults}

\author{
 \textbf{Zhengbang Yang\textsuperscript{1}},
 \textbf{Junyuan Hong\textsuperscript{2}},
 \textbf{Yijiang Pang\textsuperscript{3}},
 \textbf{Jiayu Zhou\textsuperscript{4}},
 \textbf{Zhuangdi Zhu\textsuperscript{1}}
\\
 \textsuperscript{1} George Mason University, Fairfax, VA, USA
\\
 \textsuperscript{2} University of Texas at Austin, Austin, TX, USA
\\
 \textsuperscript{3} Michigan State University, East Lansing, MI, USA
 \\
 \textsuperscript{4} University of Michigan, Ann Arbor, MI, USA
 \\
 \texttt{\{zyang30,zzhu24\}@gmu.edu}
}

\begin{document}
\maketitle
\begin{abstract}
Cognitive health in older adults presents a growing challenge. Although conversational interventions show feasibility in improving cognitive wellness, human caregiver resources remain overloaded. 
AI-based chatbots have shown promise, yet existing work is often limited to implicit strategies or heavily depends on training and label resources.  
In response, we propose a strategy-guided AI chatbot named \sysName\   that follows a dual-level conversation reasoning framework.
%
It integrates macro-level strategy planning and micro-level utterance generation to enable engaging, multi-turn dialogue tailored to older adults.
{Empirical results show that \sysName\ closely aligns with professional human caregiver behaviors in offline evaluation using real clinic data, and achieves positive user cognitive and emotional responses in interactive simulations with digital twins, which significantly outperforms AI baselines that follow implicit conversation generation.}
\end{abstract}
%
\section{Introduction} 
 
The cognitive well-being of the elderly population is a pressing social concern, as evidenced by the prevalence of cognitive disorders within this population, often exacerbated by loneliness and isolation~\cite{nicholson2012review,teo2023global}. According to WHO, approximately $14\%$ of adults aged 60 and over experience mental health disorders, projected to affect 2.1 billion individuals by 2050~\cite{aging-health}. 
Compared with other age groups, older adults are more vulnerable due to age-related changes in cognitive reserves ~\citep{salthouse2009does} and reduced social connections ~\citep{nicholson2012review, teo2023global}. 
Such impact extends beyond older individuals to families and society, resulting in a reduced life quality and increased medical burden. 
In the meantime, modest delays in cognitive decline can significantly reduce dementia prevalence, and addressing social isolation could prevent 4\% of dementia cases~\cite{livingston2020dementia}. 
Interventions through guided conversations have shown efficacy in reducing loneliness and mitigating cognitive decline  ~\citep{yu2021internet,yu2022examining,yu2023conect,fiori2012impact}. 
However, effective intervention requires sustained interaction and monitoring, which is limited by the availability of human companions, leading to inconsistent access or effectiveness. 

Advances in artificial intelligence (AI) particularly Large Language Models (LLMs) have shown promise in augmenting human expertise with conversational support.
Existing efforts range from sentiment-based  chatbots~\cite{liu-etal-2021-towards,seniortalk,ryu2020simple,tsai2021exploring,you2023beyond,meyer2020patient} to  audio assistants~\cite{yang2024talk2care,hong2024conect},
ascribed to mature text-audio transformation techniques~\cite{van2016wavenet,ren2019fastspeech}.
However, these systems primarily default to interactions with implicit goals, which may fail to drive engaging conversations that necessitate strategic multi-turn interactions tailored to older adults. 
\J{Some work has explored fine-tuning LLMs on domain-specific datasets to adapt to elderly care~\cite{sun2024enhancing}, which requires extensive labeled data and resources, while they may struggle to generalize across conversational contexts. In contrast,  we take an orthogonal approach by focusing on the inherent reasoning capabilities of LLMs through in-context learning, which offers more resource efficiency and adaptability.
}

Our goal is to provide AI-powered, \textit{\textbf{engaging} conversational support} for older adults that serves as an accessible {complement} to human companions, with the aspiration to improve their cognitive function and reduce social loneliness.
Previous clinical studies aimed at socially isolated older adults showed the existence of causal relationships between interviewer strategy and interviewee response ~\citep{cao2021causal}, 
which revealed that conversation behavior can have a \textit{measurable} influence on interlocutors, and thus inspired us to develop {principled} yet {efficient} methods that transform clinical insights into dialogue design by leveraging the advanced reasoning ability of LLMs.

In response, we propose an LLM-driven chatbot named \sysName.
%
It employs \textit{dual-layer} conversation generation that first derives categorized macro-level information to suggest meta-conversational strategies, which then guides the micro-level utterance generation to improve both user engagement and cognitive outcomes over multi-turn dialogue interactions.
{\sysName\ is evaluated using both real, de-identified real-world dialogues between older adults and professional human caregivers of a clinical trial \citep{iconect-detail}, and synthetic interactive conversations using simulated users (\ie\ digital twins) ~\citep{hong2024conect} modeled on such data.} 

\begin{figure}
    \centering
    \includegraphics[width=0.6\linewidth]{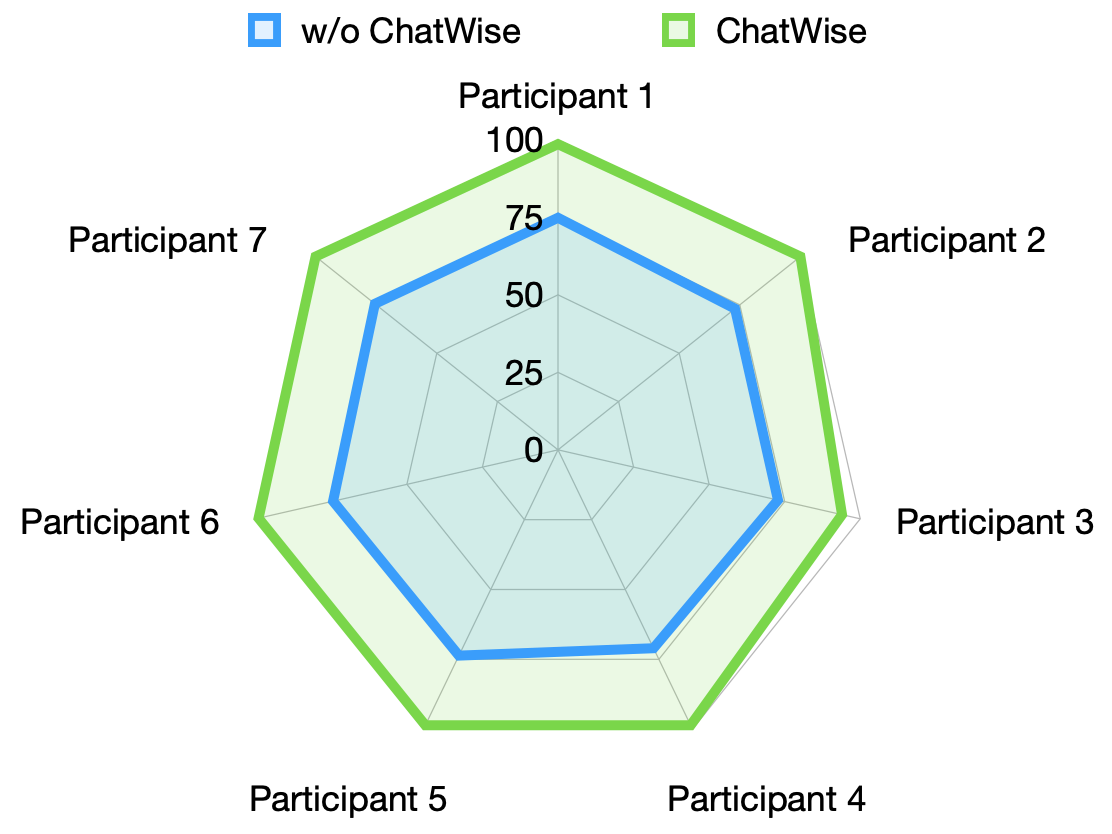}
    \centering
    \caption{ Alignment between conversational strategies of AI chatbots and human professionals across diverse participant data. Our design shows closer alignment with caregiver behavior in real conversation contexts compared to baseline chatbots. See Sec~\ref{sec:evaluation} for full results.}
    %
    \label{fig:overall}
\vspace{-0.2in}
\end{figure}

Our work provides multifold contributions:
    (i) We introduce a clinically grounded chatbot that highly aligns with the behavior of professional human caregivers to older adults (Sec \ref{sec:eval-real}). 
    Figure \ref{fig:overall} overviews its alignment performance.
    It also empirically enhances simulated users' engagement and cognitive status, which significantly outperforms baseline chatbots~(Sec \ref{sec:eval-synthetic}). 
    (ii)  \sysName\ follows a tuning-free, in-context learning framework for daily conversational support. 
    Comparative studies demonstrated that providing macro-level strategies to guide conversation generation is the key contributor to enhancing user engagement. 
    (iii)  
    %
    Its dual-level policy design can be readily applied to various LLMs, with consistent user cognitive gains across different backbone LLMs.
    %
    
    %
     

\section{Related Work}
\paragraph {AI-powered Chatbots for Older Adults} have shown feasibility in preventing or detecting the cognitive issues to assist human professionals.
Recent efforts span commercial products \citep{seniortalk,elliq} and research prototypes focused on emotional support and audio assistance.
%
\citep{sun2024enhancing} performed supervised fine-tuning on LLMs to enhance their performance on specialized nursing and elderly care tasks.
\Zhengbang{~\citet{yang2024talk2care} introduced an LLM-based voice assistant designed to bridge communication between older adults and their healthcare providers. \citet{liu-etal-2021-towards} built an annotated dataset to tune LLMs for emotional support tasks. \citet{ryu2020simple} developed a chatbot for the mental health of the elderly. \citet{hong2024conect} constructed digital twins of the elderly.}
Unlike prior work, our approach prioritizes user conversational \textit{engagement} through in-context reasoning, which has been empirically shown to enhance user cognitive status.
\paragraph{Dialogue Systems for Mental Health or Cognitive Stimulation: }
%
 
Recent studies have leveraged LLMs for the augmentation of emotional support conversations through \textit{generated} dialogue data ~\citep{zheng-etal-2023-augesc, jiang-etal-2023-cognitive}.  ~\citet{10.1145/3613904.3642538} assessed the acceptability of GPT-4-based conversational agents among people with dementia (PwD), and highlighted the importance of design considerations for this sensitive population. 
Similarly, ~\citet{10508610} implemented a conversational robot embedded with ChatGPTs for reminiscence therapy, with authors as simulated users, which showcased the potential of generative AI to support cognitive stimulation through interactive storytelling. 
Distinct from these works, our research introduces a dual-level dialogue approach combined with comprehensive evaluation methodologies to provide a finer-grained understanding of conversational engagement.


\paragraph{Conversational Strategies: }
\citet{liu-etal-2021-towards} curated a dataset with annotated strategies, demonstrating the effectiveness of \textit{Helping Skills Theory}~\cite{hill2020helping} in providing emotional support. 
\citet{YUAN2023100384} examined the causal relationships between dialogue acts (DAs) and participants’ emotional states in a clinical trial \citep{iconect-detail}, emphasizing the impact of strategic interventions in tele-mediated dialogues.
\citet{seo2021learning} identified key strategies for improving child patient-provider communication through semi-structured interviews. 
\Zhengbang{However,} few works have systematically integrated these strategies into the reasoning flow of AI chatbots. Our approach fills this gap by contextualizing structured conversational strategies for automatically enhancing user engagement.

\paragraph{Multi-Turn Chatbot Exploration: } Recent advances in AI dialogue systems predominantly focused on short-turn interactions 
~\citep{owan2023conversational,dam2024complete,gao2025regressing}, with relatively limited attention to the challenges of multi-turn exchanges. 
While some pioneering research has explored multi-turn optimization through Reinforcement Learning (RL) approaches for LLMs \citep{verma2022chai,zhou2024archer,abdulhai2023lmrl,gao2025regressing}, these methods were not specifically designed for supporting senior dialogue engagement.
Our inference-based method introduces a strategy-compatible framework that aligns with the principles of RL-driven optimization,
which can enable future extension to further enhance multi-turn conversation engagement.

\begin{figure*}[htbp!]
    \centering
    \includegraphics[width=0.8\linewidth]{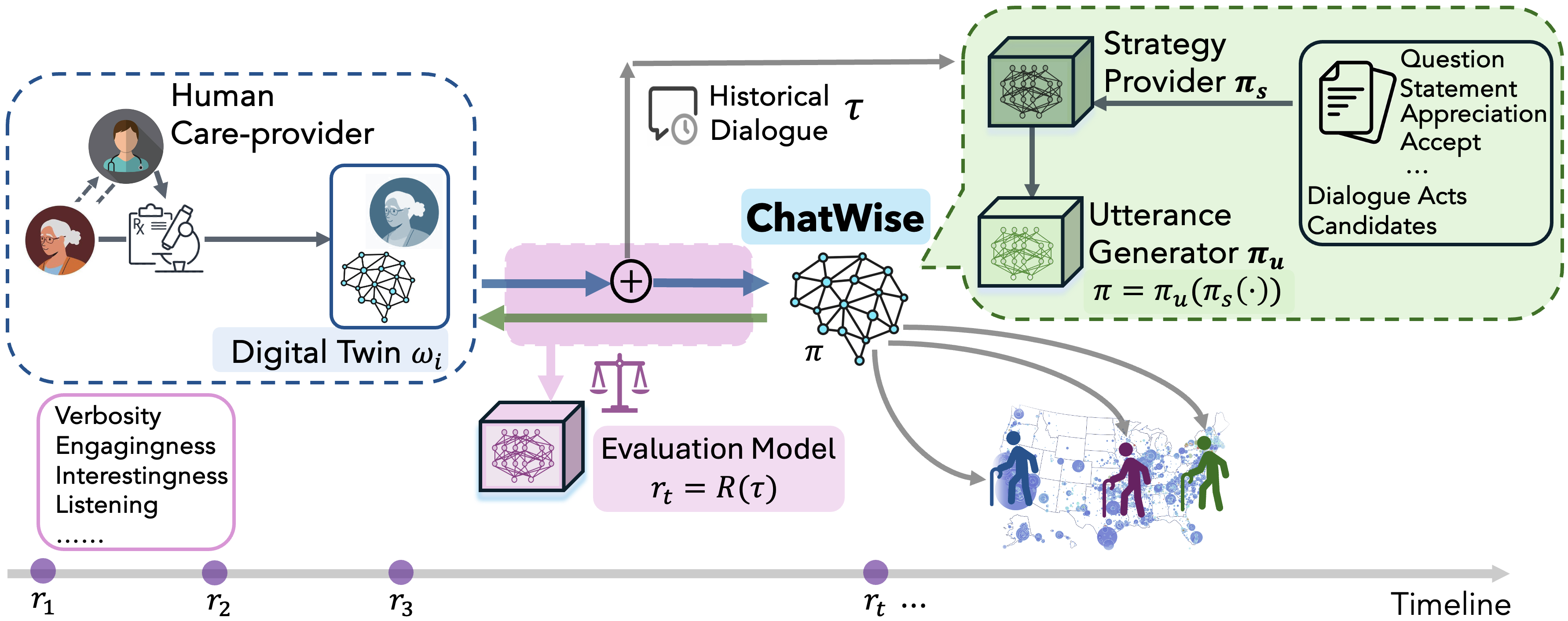}
    \caption{ \sysName\  employs a dual-level policy design that contextualizes a clinically-derived \textit{strategy pool} to generate \textit{macro-level} actions, which then guide \textit{micro-level} utterance generation. 
    It is iteratively developed using digital twins as simulated users.
    Prior to a future real user study, real and simulated data analysis show that \sysName\ closely aligns with professional human caregivers' behavior, leading to cognitive gains of simulated users accumulated over multi-turn interactions.}
    \label{fig:ChatWise}
\end{figure*}

\section{\sysName\ Design} \label{sec:sys-design} 

\sysName\ encompasses a hierarchical framework that integrates the insights of traditional clinical studies to drive LLM reasoning.
%
It features a dual-level design with a \textit{strategy provider} $\pi_s$ that generates macro-level conversation strategies, which guides a \textit{utterance generator} $\pi_u$ to produce micro-level responses. 
In this paper, we interchangeably refer to \sysName\ as the \textit{interviewer} or the \textit{moderator},  a role commonly defined in communication studies that guides and facilitates the conversation ~\citep{Taboada2006},

For clarity, we denote a multi-turn conversation as $\tau=\{\mu_0, u_0, x_1, u_1, x_2, u_2, \cdots\}$ which comprises  interviewee (user) utterances $x_t$ and interviewer (\sysName) utterances $u_t$ at time step $t$, with a conversation starting point $\mu_0$.
$s_t=\{\mu_0,u_0,\cdots,x_t\}$ denotes the historical conversation up to timestep $t$.
We then formulate  \sysName\ as a dual policy $\pi=\{\pi_s,\pi_u \}$,
which maps the conversational \textit{state} $s_t$ to the next utterance $u_t$ provided to the user, conditioned on meta information $z$ given to the strategy provider $\pi_s$:
$$u_t \sim \pi(\cdot|s_t,z) \equiv \pi_{u}( \cdot| \pi_{s}(s_t|z)).$$ 
Below, we illustrate the process of this hierarchical conversation generation.

\subsection{Strategy Pool}\label{sec:strategy-pool}
 To first develop clinically grounded strategies, we establish a strategy pool containing dialogue acts (DA) curated from clinical intervention studies involving older adults.
These DAs, such as \textit{asking an open question} (\eg\ ``What is your favorite movie'') or  \textit{showing empathy} (\eg ``That sounds really meaningful''), serve as  macro-level actions and atomic communicative units that convey distinct conversational intentions~\cite{searle1976classification}, 
which have also shown causal relationships with participant emotions in conversations~\cite{cao2021causal}. 
Specifically, we extract DAs that exist in conversations from a real tele-health clinical trial~\cite {YUAN2023100384}, which represents a subset of the complete DAMSL taxonomy~\cite{allen1997draft,searle1976classification}.
We further augment this DA with strategies from prior emotional support dialogue dataset~\citep{liu-etal-2021-towards} to form a comprehensive strategy pool $\mathcal{A}$. 
We then treat each strategy serves as a macro-level action candidate $a \in \mathcal{A}$ and refine it with a \textit{definition} and an \textit{example} to serve as in-context information $z$ for enhancing the reasoning of strategy $\pi_s$
(See {Appendix} \ref{appendix:DA} for a complete DA set).

\subsection{Emotion Annotation}  
To enhance context-awareness during conversations and support post-conversation analysis, the strategy provider $\pi_s$ is additionally prompted to infer the user's current emotion based on historical conversations $s_t$. 
At timestep $t$ of the conversation, $\pi_s$ will categorize user emotion into one of the five classes: \textit{joy, neutral, sadness, anger, and surprise}, given a calibrated system prompt. 
Our comparative study shows that \sysName\ can lead to detectable positive emotion changes over multi-turn conversations, which aligns with conversations provided by human professionals~\J{(Sec~\ref{sec:user-emotion})}.

\subsection{From Strategy to Utterance Generation}
The strategy provider $\pi_s$ processes the dialogue history $s_t$, contextualized by the strategy pool information $z$ to decide the most appropriate strategies for continuing a conversation.
We limit each interviewer's utterance to contain one or two strategies, \ie\ $\mathbf{a}_t =\{a_{t}^i\}_{i\leq2} \sim \pi_s(\cdot|s_t,z)$.    
Our rationale follows counseling studies~\citep{zhang2020balancing}  that dialogue intentions can be \textit{forward}, \eg\  initiating new topics via an open question, or \textit{backward}, \eg\  responding with acknowledgment, or both to transit between intentions or topics.
Thus,  $\pi_s$ is prompted to comply with one of the three conditions: (i) a forward strategy, (ii) a backward strategy, or (iii) a backward strategy followed by a forward strategy.

The selected strategies $\mathbf{a}_t$, and \textit{optionally}, the user’s current emotion label $e_t$, and the conversation history $s_t$ are used as inputs to the utterance generator $\pi_u$. 
Our ablation study showed that providing meta-level strategies has a dominant effect over providing emotion information alone for conversational engagement (Sec \ref{sec:ablation}).
To ensure natural flow, $\pi_u$ will first improvise a few rounds as warm-ups before adopting the suggested strategies. 
We also draw on clinical guidance to let $\pi_u$ encourage users to choose topics rather than imposing them.
Figure~\ref{fig:ChatWise} overviews this design.

\section{Experiments}  \label{sec:evaluation}
We conducted various experiments focusing on answering the following  questions: 

\noindent \textbf{Q1.} Does \sysName's behavior align with human professionals when responding to real conversation context sampled from clinic trials?\\
\noindent \textbf{Q2.} How does \sysName\ compare to baseline AI alternatives in terms of multifaceted conversational engagement metrics?\\
\noindent \textbf{Q3.} How do factors such as the conversation turns and different user characteristics influence \sysName's performance? 

To address these questions, we evaluated \sysName\ on two complementary scenarios: we first conducted \textit{\textbf{offline}} evaluations using data from a real clinical trial  (Sec~\ref{sec:eval-real}), then we performed \textit{\textbf{interactive}} experiments, in which conversations are stochastically generated between \sysName\ and digital twins as simulated users (Sec~\ref{sec:eval-synthetic}).
Below, we summarize the configurations, metrics, and main results of each experimental setting and defer more details to the Appendix. 

\subsection{Offline Evaluation on Real Clinical Data}\label{sec:eval-real}

\subsubsection{Data Preprocessing}
We extracted dialogues from the I-CONECT clinic trials~\cite{iconect-detail}, with each in text format converted from video conversations. 
Short dialogues below  {40} \textit{turns} were excluded, where a \textit{turn} represents an uninterrupted, continuous utterance by a single speaker.
 Two dialogue sets were subsampled and de-identified:
(1) 150 randomly sampled dialogues  to assess the overall strategy alignment,
and (2) one dialogue per week from 7 randomly selected participants to assess \sysName's alignment robustness to different participants and time periods over the clinical trial. 

\subsubsection{Offline Evaluation Metric} 
We define Strategy Match Percentage (SMP) to quantify the alignment between the conversational strategies generated by a chatbot and those generated by a human professional caregiver from the I-CONECT clinic data.
Let $\mathbf{a}_t$ denote the set of caregiver-provided strategies annotated from a real conversation $\tau$ at turn (timestep) $t$, 
and $\mathbf{a}'_t$ the strategies from a simulated moderator (\sysName\ or utterance generator), conditioned on the same real context $s_t$.
%
Let $\mathbb{I}(a'_t \in \mathbf{a}_t  )$ be an indicator function that returns 1 if a chatbot-generated strategy $a'_t$ matches one of the human strategies $\mathbf{a_t}$, and 0 otherwise. The SMP is computed as following:

$$\text{SMP}_t(\mathbf{a}_t, \mathbf{a}'_t) =\frac{1}{|\mathbf{a}_t|}\sum_{a'_t \in \mathbf{a}'_t}\mathbb{I}(a'_t \in \mathbf{a}_t).$$

\begin{figure}[htbp!]
    \centering
    \includegraphics[width=1\linewidth]{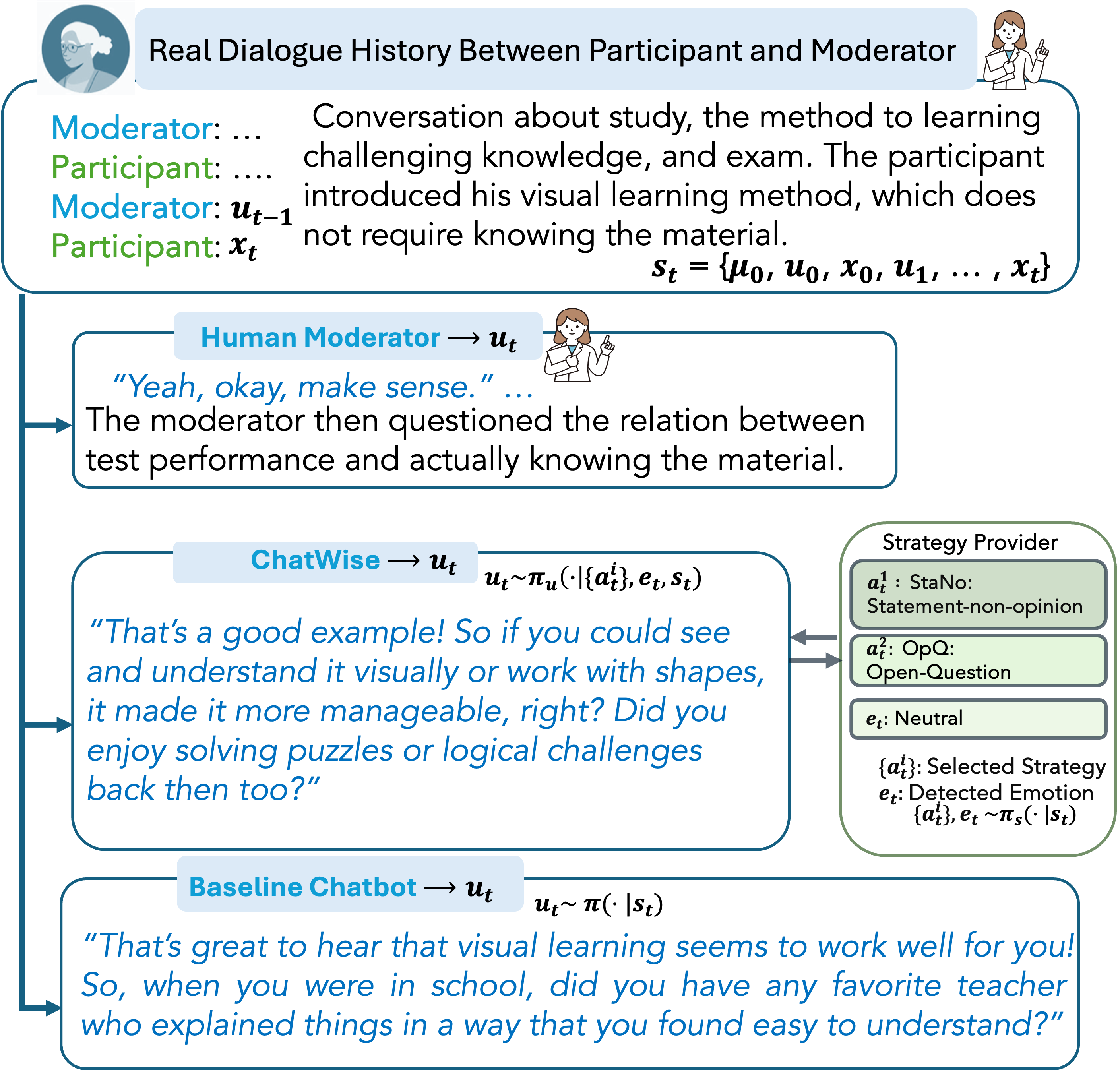}
    \caption{Comparison of moderator strategies: Given real dialogue contexts, responses from a baseline (generated), \sysName\ (generated), and the human moderator (real) are compared \textit{w.r.t.} underlying strategies. A real sample is shown here while omitting participant data.}
    \label{fig:dialogue-example}
\end{figure}

\subsubsection{Strategy Label Annotation}

Given a real dialogue denoting as $\tau=\{\mu_0,u_0, x_1,u_1, x_2,u_2,\cdots\}$ with caregiver (moderator) utterance $u_t$ and participant utterance $x_t$ at turn $t$,  
we leveraged the macro-level strategy definition in Sec~\ref{sec:strategy-pool} to prompt an LLM as annotator (GPT-4o) and detect strategies  $\mathbf{a}_t$ behind  $u_t$ as the golden label. 
Denoting $s_t=\{\mu_0,u_0, x_1, \cdots,x_t\}$ as the real conversation context, we feed $s_t$  to a \textbf{\textit{baseline}} chatbot to get utterance action $\tilde{u}_t$. The baseline is identical to the utterance generator $\pi_u$ of \sysName\ yet does not receive strategy guidance from $\pi_s$. Similarly, we get strategy labels $\tilde{\mathbf{a}}_t$ behind $\tilde{u}_t$ using the above annotator.
We then input $s_t$  to \sysName's strategy provider $\pi_s$ to get proposed strategy $\mathbf{a}'_t \sim \pi_s(s_t)$.
We report the strategy alignment of $\text{SMP}_t(\mathbf{a}_t, \mathbf{a}'_t)$ (w/ \sysName) and $\text{SMP}_t(\mathbf{a}_t, \tilde{\mathbf{a}}_t)$ (w/o \sysName) over different turns $t$ and average the result across dialogues. Figure \ref{fig:dialogue-example} illustrates the process of collecting each moderator's response based on real conversation history.

\subsubsection{Overall Strategy Alignment}

\begin{figure}[!htbp]
    \centering
    \includegraphics[width=0.8\linewidth]{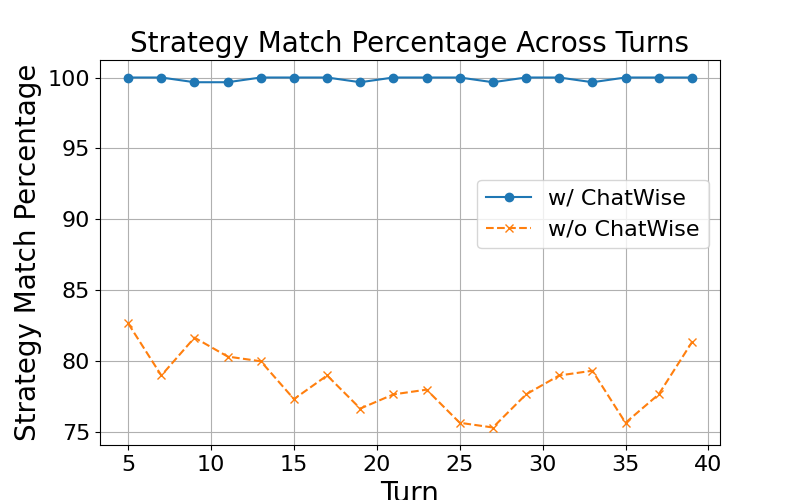}
    \centering
    \caption{SMP between human caregiver and chatbots on real conversation data, where w/o \sysName\ denotes the baseline without strategy guidance. Our method strongly aligns with human caregiver behavior.}
    \label{fig:Overall_Strategy_Alignment}
    \vspace{-0.2in}
\end{figure}

As shown in Figure \ref{fig:Overall_Strategy_Alignment}, which averages results over 150 dialogues, \sysName\ consistently achieves an SMP close to 1.0, indicating a strong alignment with the strategies employed by real caregivers throughout the conversation. 
In contrast, the baseline chatbot deviates from human behavior when lacking action guidance. 
Both chatbots used GPT-4o  as the utterance generator $\pi_u$, and o3-mini was used for $\pi_s$ in \sysName. 
We do not employ micro-level utterance similarity, as the I-CONECT interventions involve spontaneous daily conversations, rather than fixed question answering, unlike domains in programming or math.
Whereas, on a macro level, \sysName\ mostly mirrors the strategic choices of human professionals who are well trained to follow clinical protocols.

\begin{figure}[htbp!]
    \centering
    \includegraphics[width=0.7\linewidth]{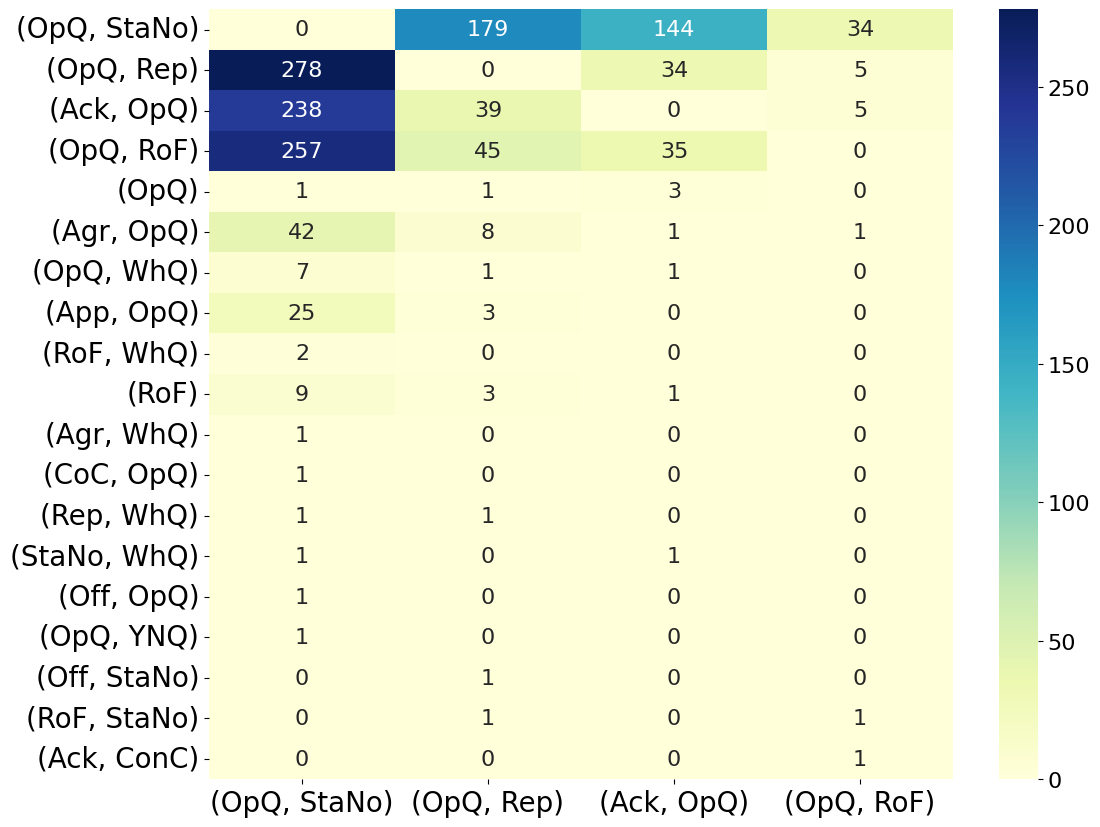}
    \caption{ Strategy distribution for dialogue samples where $\mathbf{a_t} = \mathbf{a}'_t  \neq \tilde{\mathbf{a}_t}.$ The vertical axis indicates ground-truth strategies $\mathbf{a_t}$ extracted from human professional utterances, and the horizontal axis denotes strategies $\tilde{\mathbf{a}}_t$ detected from the baseline chatbot. } 
    \label{fig:strategy-discrepancy}
    \vspace{-0.2in}
\end{figure}

To illustrate how our method behaves differently from the baseline, we collect the real dialogue context and different moderator responses where 
$\mathbf{a_t} = \mathbf{a}'_t  \neq \tilde{\mathbf{a}_t}.$
and show the discrepancy heatmap. 
Raw dialogue content is omitted to comply with the data usage agreement.
\J{Results in Figure~\ref{fig:strategy-discrepancy} demonstrate that while ``\textit{open question}'' is a commonly dominant strategy (forward DA) for both human and chatbot moderators, the primary strategy difference lies in how they lead into the question (\eg\ ``\textit{acknowledge}'', ``\textit{restatement}''), which may in turn influence the specific question ultimately asked.}

\begin{figure}[!htbp]
\vspace{-0.05in}
    \centering
    \includegraphics[width=0.95\linewidth]{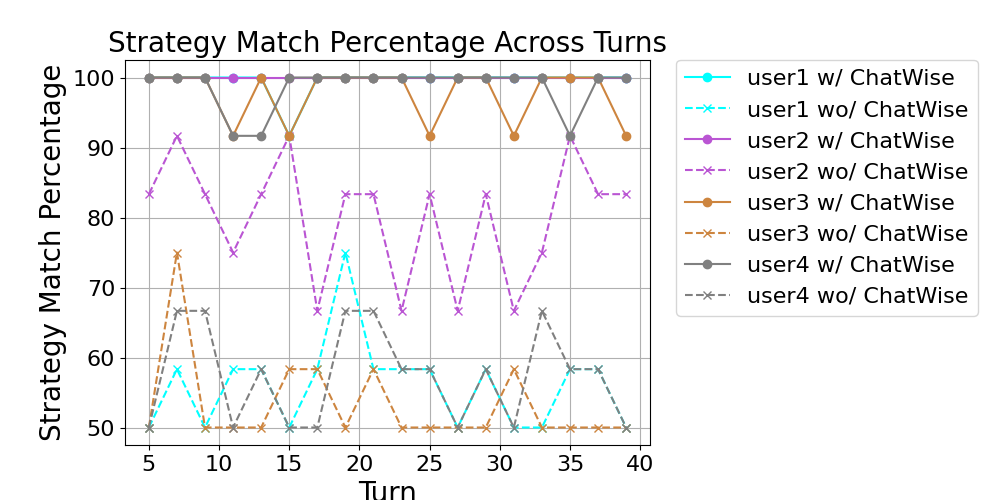}
    \centering
    \caption{SMP broken down by participants over 40 conversation turns, where each color indicates a specific person. Solid lines represent \sysName, and dashed lines denote the baseline. \sysName\ consistently aligns to human caregivers given different participant contexts.}
    \label{fig:I-CONECT_user}
\vspace{-0.15in}
\end{figure}

\paragraph{Robustness to Participant Heterogeneity:}
Figure \ref{fig:I-CONECT_user} presents the SMP of \sysName\ across dialogue turns among participants. 
\sysName\ consistently outperforms the baseline and maintains notably higher SMP. 
These results demonstrate the robust alignment of \sysName\  with human caregiver strategies across diverse participants and their interaction contexts, which implies its potential for personalized dialogue systems.

\begin{figure}[!htbp]
    \centering
    \includegraphics[width=0.8\linewidth]{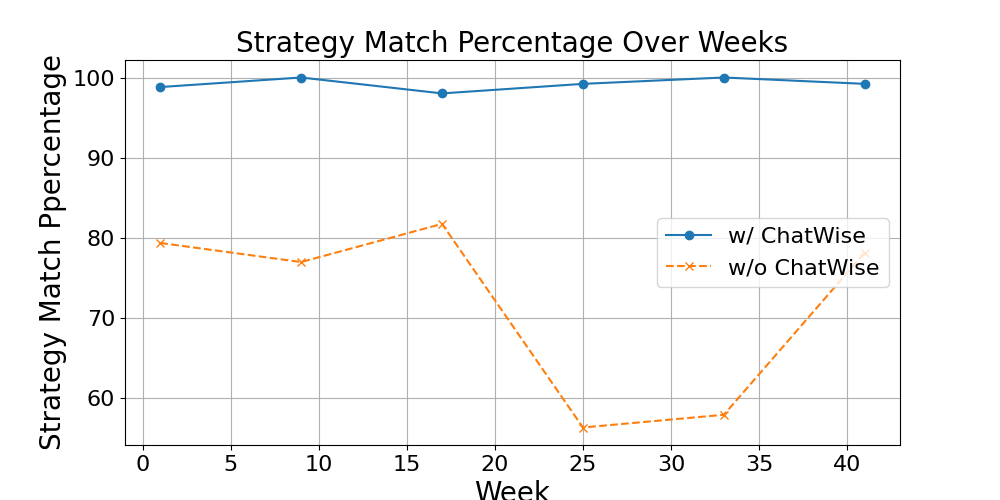}
    \centering
    \caption{Strategy alignment of \sysName\ is consistent as real conversations progress over weeks.}
    \label{fig:I-CONECT_time}
    \vspace{-0.2in}
\end{figure}

\paragraph{Alignment Consistency Over Timeline:}
Figure \ref{fig:I-CONECT_time} presents the SMP over time, where dialogues were sampled from different weeks over the clinical intervention.
\sysName\ maintains consistently high SMP (near 1.0) across all timesteps, where the baseline shows lower and unstable performance.
This indicates \sysName's temporal robustness and its potential to facilitate the development of long-term cognitively supportive dialogues.

\vspace{-0.1in}
%


\subsection{Evaluation with Digital Twins}\label{sec:eval-synthetic}

\subsubsection{Conversation Generation}
\label{sec:Conversation_Generation}

While evaluation on real clinic data shows strong alignment between \sysName\ and professional caregivers, the conversations are fixed, thus lacking indications whether and how \sysName\ may influence user cognitive or emotional status during conversations.
To enable controlled and responsive development of \sysName\ prior to future human user studies, we used 9 digital twins provided by ~\citet{hong2024conect} as simulated users with different personas, which are fine-tuned LLMs calibrated to mimic the linguistic behaviors of older adults, some of whom show MCI symptoms. All digital twins are trained using \textit{real, de-identified} dialogue data from the I-CONECT clinical trial. 
We collected 20 trajectories of conversations between \sysName\ (moderator) and each digital twin as a user, with each conversation containing 20 turns. The same settings were applied to the baseline chatbot for fair comparisons.


\subsubsection{Metrics for Interactive Engagement}

\noindent\textbf{User Verbosity:}
Given a conversation sequence $\tau$ that consists of user utterance $x$ and moderator utterance $u$, we primarily measure user engagement through the user's talkativeness compared to the moderator, \ie\ the user's verbosity ($v$),  defined as 
$v=\frac{\sum_{x \in \tau} |x|}{\sum_{u \in \tau}|u|},$ where $|x|$ ($|u|$) denotes the number of tokens in an utterance $x$ ($u$).
This metric follows a clinical study~\citep{yu2021internet} that suggests reducing the moderator talkativeness while encouraging participant (user) expression. 
\vspace{-0.1in}
\paragraph{Cognitive Win Rate: }

\citet{see-etal-2019-makes} defined different aspects for evaluating conversation quality, from which we select \J{3} that are focused on assessing user cognitive status: {\textit{Listening}}, {\textit{Fluency}}, and {\textit{Making Sense}}. 
All three metrics evaluate the \textit{\textbf{user}} behavior rather than the overall dialogue quality considering both user and moderator behaviors.  

Instead of providing a numerical score for each cognitive metric that might be unstable, we adopt the \textit{\textbf{Win Rate (WR)}} definition from prior work ~\citep{NEURIPS2023_a85b405e} to compare pairs of dialogues generated with and without \sysName.
Dialogue pairs are randomly matched so long as they are collected from the same digital twin (simulated user) for evaluation.
WR is defined as the proportion of pairs in which the dialogue generated with \sysName\ is preferred over the baseline using an LLM-as-judge, based on a given cognitive metric.

To mitigate the LLM's \textit{position bias} issues observed in prior work~\cite{shi2024judging,yu2024mitigate}, we excluded sample pairs with inconsistent preference labels when the order of the two samples was reversed and calculated WRs using only the remaining consistent pairs.

\begin{table}[htbp!]
    \centering
    \resizebox{0.5\textwidth}{!}{%
        \begin{tabular}{lcccc}
            \toprule
            \textbf{Strategy Provider} & \textbf{Verbosity } $\uparrow$& 
            \textbf{Listening} $\uparrow$ & \textbf{Fluency} $\uparrow$ & \textbf{Making Sense} $\uparrow$ \\
            \midrule
            baseline  & 0.7398 & \textbf{0.5249} & 0.4986 & 0.5024 \\ 
            GPT-4o        & 0.8635 & 0.4962 & 0.4748 & 0.4786 \\
            o3-mini      & \textbf{0.8643}  & 0.4884 & \textbf{0.5368} & \textbf{0.5180} \\
            Llama 3.1-405B & 0.8083 & 0.4407 & 0.4926 & 0.4963  \\
            \bottomrule
        \end{tabular}
    } 
    \caption{Multifaceted evaluations on cognitive engagement with data sampled by interacting with digital twins over 20 turns, using different LLMs serving as strategy providers,  where \textit{Listening}, \textit{Fluency}, and \textit{Making Sense} metrics report win rates, and baseline denotes a conversation generator without receiving strategy guidance.  All evaluations focus on \textit{user} reactions. 
    }
    \label{tab:overall_performance}
    \vspace{-0.2in}
\end{table}

\subsubsection{Models and Baseline}
We evaluated \sysName\ with 3 different LLM backbones as strategy providers: GPT-4o, o3-mini, and Llama3.1-405B. 
For each setting, a baseline adopts the same utterance generator as \sysName\ while without receiving strategy guidance. 
The WR of \textit{baseline}  is reported as the  complement of \sysName's average WR: $\text{WR}_\text{baseline} = 1 - \text{average}(\text{WR}_{GPT-4o} + \text{WR}_{o3-mini} + \text{WR}_{Llama3.1-405B}),$  where $\text{WR}_{X}$ indicates the WR of \sysName, given model $X$ as the strategy provider.

\subsubsection{Performance Overview}
\paragraph{Effectiveness of Strategy-Guided Generation:} As shown in Table \ref{tab:overall_performance},  our design mostly demonstrated better user engagement compared to baselines across  tested LLMs as the strategy provider.  
Particularly, the best-performing model in our setting, o3-mini, was optimized for STEM reasoning tasks, which indicates that strong reasoning ability can also benefit inferring dialogue strategies.

\begin{figure}[!htbp]
    \centering 
    \begin{subfigure}[t]{0.49\linewidth}
        \includegraphics[width=\linewidth]{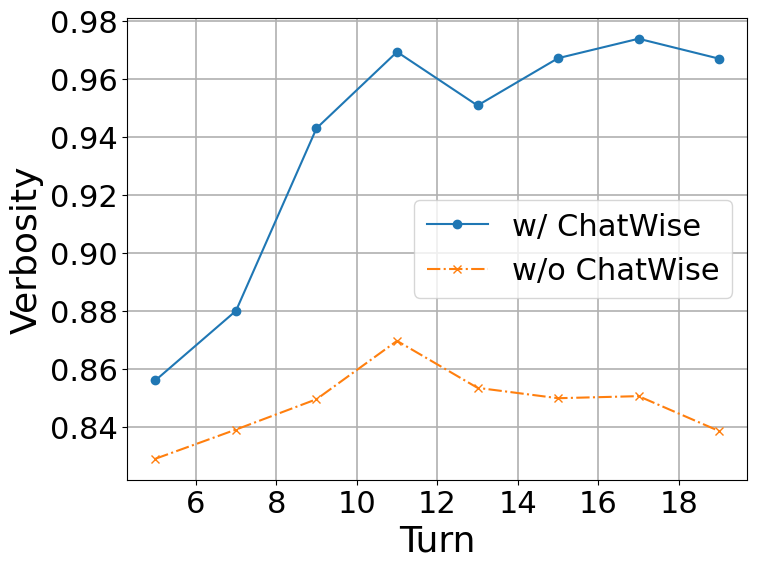}
        \caption{User Verbosity.}
    \end{subfigure} \hfill
    \begin{subfigure}[t]{0.49\linewidth}
        \includegraphics[width=\linewidth]{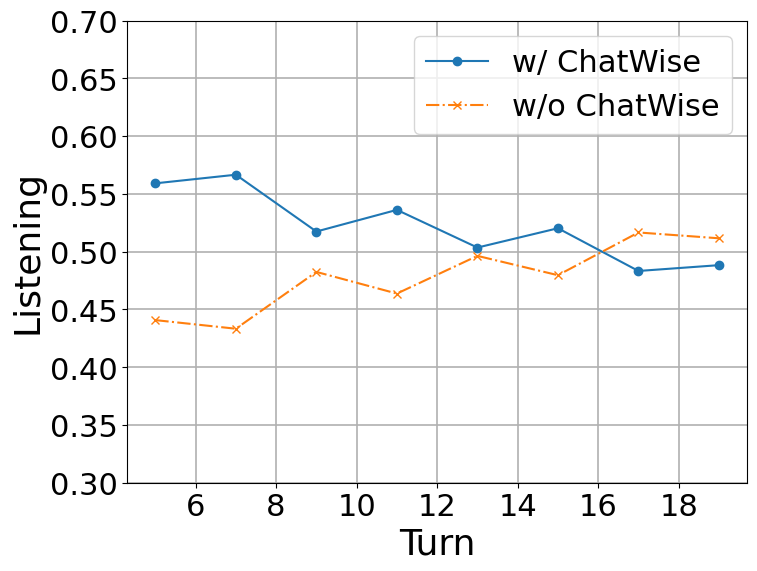}
        \caption{WR: Listening.}
        \captionsetup{justification=centering}
    \end{subfigure}
    \begin{subfigure}[t]{0.49\linewidth}
        \includegraphics[width=\linewidth]{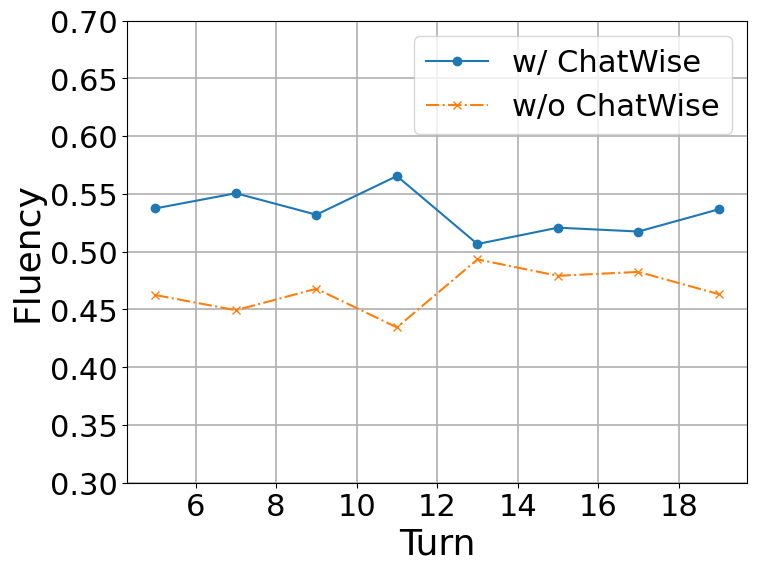}
        \caption{WR: User Fluency.}
    \end{subfigure} \hfill
    \begin{subfigure}[t]{0.49\linewidth}
        \includegraphics[width=\linewidth]{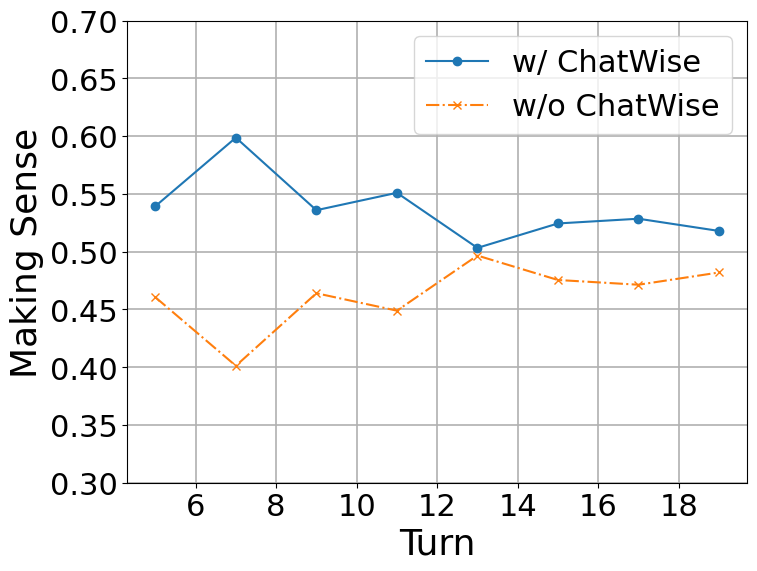}
        \caption{WR: Making sense.}
        \captionsetup{justification=centering}
    \end{subfigure}
    %
    \caption{Multifaceted Engagement Evaluation across conversation turns, excluding the first two warm-up turns. \sysName's performance gain over the baseline exists over multi-turn dialogues.
    }
    \label{fig:accumulated}
\vspace{-0.2in}
\end{figure}

\paragraph{Performance with Dialogue Progression:} 
We analyzed \sysName's performance over increasing dialogue length by truncating conversations at turn $t$ and computing metrics accordingly, as shown in Figure \ref{fig:accumulated}. 
\J{Results revealed a clear upward trend in \textit{verbosity}, which indicates increased talkativeness of the user rather than the \sysName, as opposed to the baseline.
Meanwhile, the WR gains of \sysName\ either remain stable or show a slight decline as conversations progress, which we infer are subject to the length bias of LLMs~\cite{dubois2024length} as the overall number of tokens of \sysName-generated dialogues tends to decrease in later turns. 
}

%

\vspace{-0.1in} 
\subsection{Ablation Study} \label{sec:ablation}
%
We evaluated \sysName\ with and without user \textit{emotion} in the strategy provider's output.  
As shown in Table \ref{tab:ablation}, where both strategy providers used GPT-4o as the backbone, removing emotion information led to only a slight drop of performance, while removing \sysName\ entirely resulted in a much lower \J{verbosity score}, which highlights that strategy guidance itself is the key driver of user engagement. 

\begin{table}[h]
    \centering
    \resizebox{0.49\textwidth}{!}{%
        \begin{tabular}{lccc}
            \toprule
            \textbf{Method} & \textbf{\sysName} & \textbf{\sysName\ w/o emotion} & \textbf{w/o \sysName} \\
            \midrule
            \textbf{Verbosity} & 0.8635 & 0.8463 & 0.7398 \\
            \bottomrule
        \end{tabular}
    }
    \caption{Ablative study on the performance of user verbosity when removing different contextualized information. w/o emotion denotes the strategy provider without user emotion information as input.}
    \label{tab:ablation} 
\vspace{-0.2in}
\end{table}

\subsection{\sysName's Robustness to User Persona}
\begin{figure}[!htbp]
    \centering
    \includegraphics[width=0.5\linewidth]{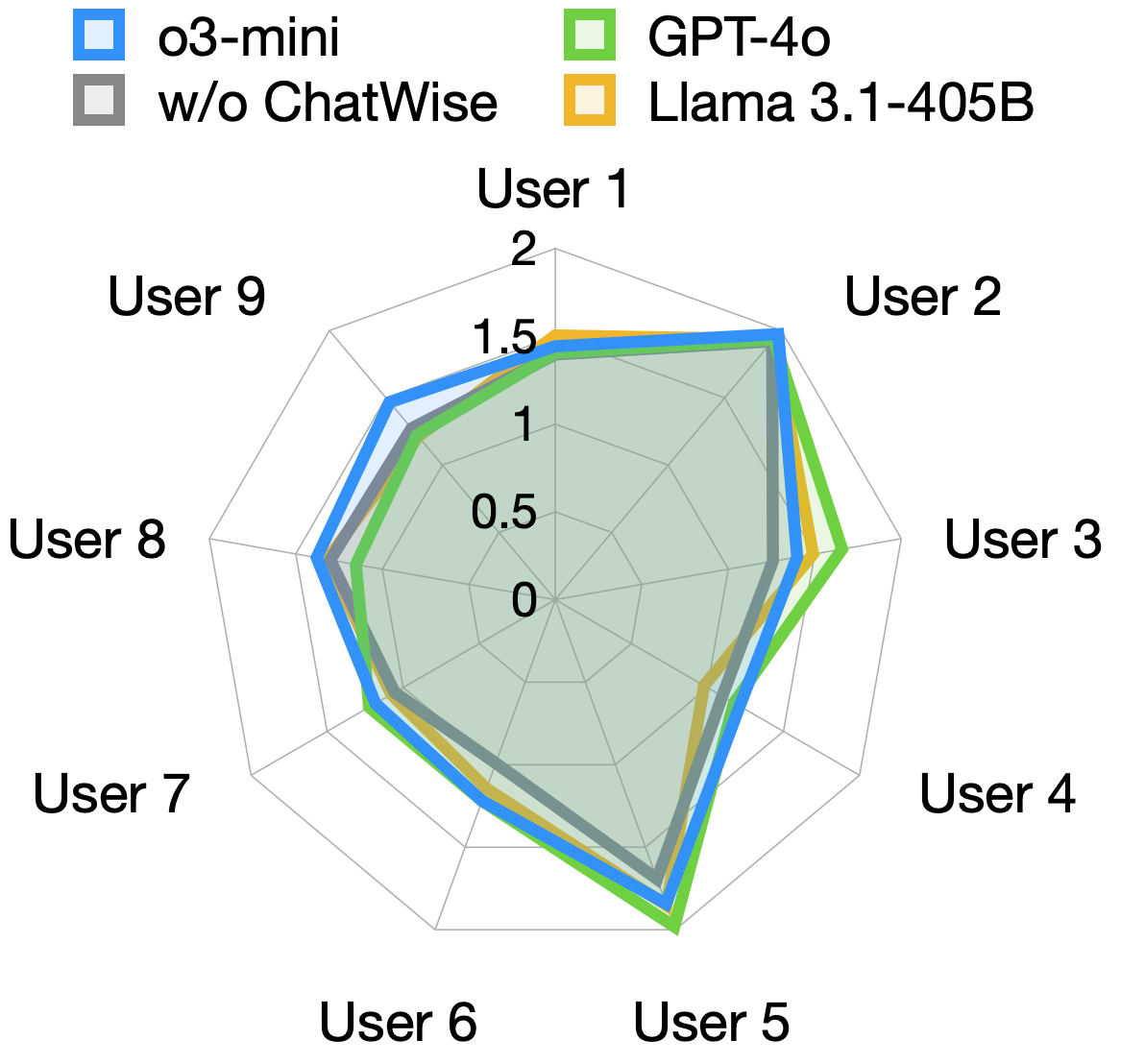}
    
    \caption{User \textit{verbosity} log-normalized across 9 digital twins. Given varying strategy provider backbones, our design consistently outperforms the baseline (w/o \sysName) with non-strategic conversations.}
    \label{fig:robustness} 
    \vspace{-0.15in}
\end{figure}

%
%
%
As shown in Figure \ref{fig:robustness}, the gain of \sysName\ interacting with  9 simulated users is consistently significant, in which the metrics were {log-normalized.}
This indicates the robustness of \sysName\  given varying senior characteristics.
 
\subsection{Analysis of User Emotion Transitions} \label{sec:user-emotion}
%
To analyze if \sysName\ can support positive emotional shifts during conversations, we selected four digital twins and collected 40 additional dialogues from each for deeper analysis, with the main findings summarized below.

\begin{figure}[!htbp]
    \centering
    \includegraphics[width=\linewidth]{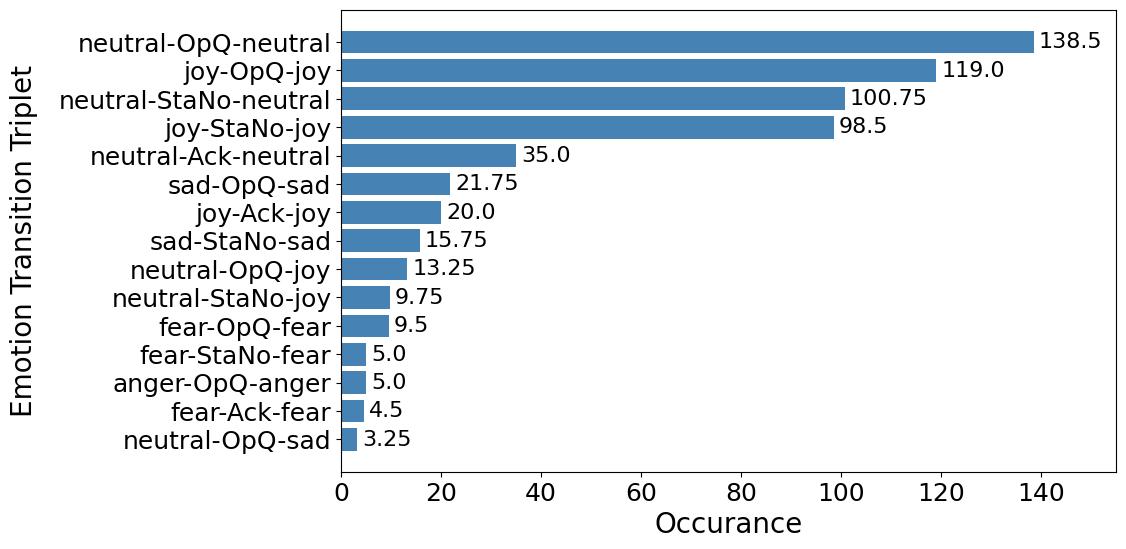}
    \caption{Average occurrence of each emotion transition triplet across the samples of each digital twin.}
    \label{fig:insight_DA_emo} 
\end{figure}

\paragraph{Transient user emotions:}\label{sec:user_emotion}
We define an emotion transition triplet as $(\text{e}_t, u_t, \text{e}_{t+1})$, where $\text{e}_t$ ($\text{e}_{t+1}$) represents the user's emotion at turn $t$ ($t+1$), and $u_t$ is the moderator utterance in between.
We computed the average occurrence of emotion triplets and showed the top 15 most frequent in Figure \ref{fig:insight_DA_emo}. Most triplets reflect unchanged user emotions, which indicates difficulty in either influencing or detecting emotional shifts within a short turn.

\begin{figure}[ht!]
    \centering
    \begin{subfigure}{0.85\linewidth}
        \includegraphics[width=\linewidth]{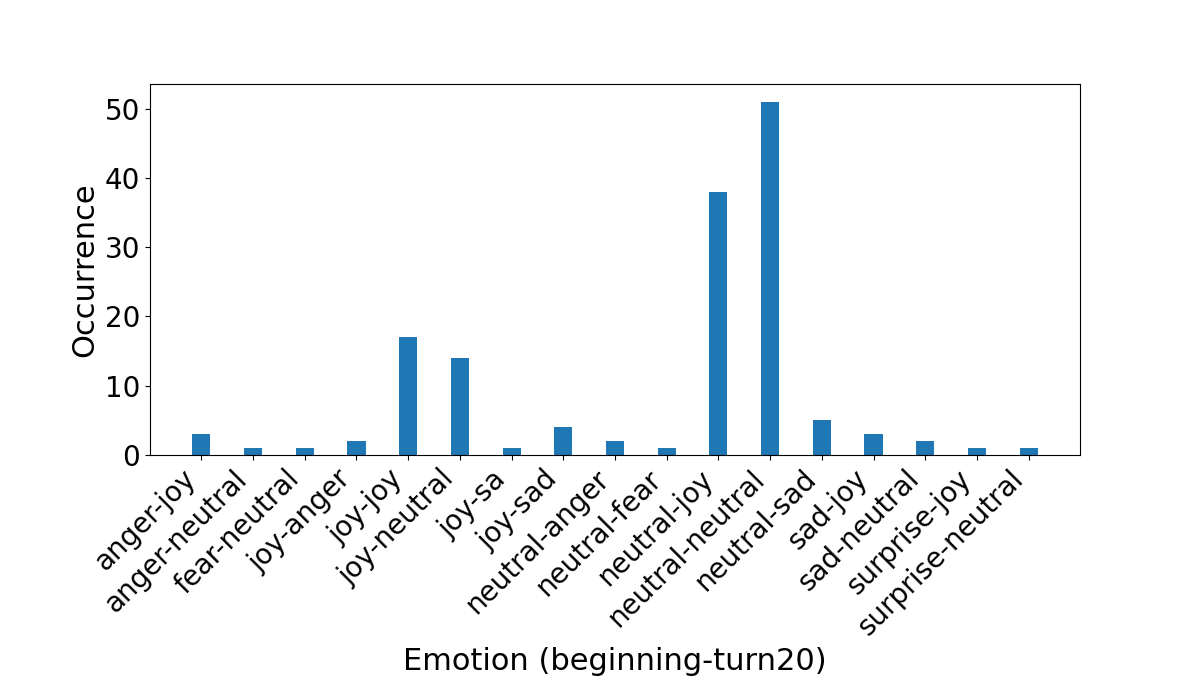}
        \captionsetup{justification=centering}
        \vspace{-0.15in} 
        \caption{\textbf{\textit{Real user}} emotion transitions summarized from 150 dialogues in the I-CONECT study. \label{fig:I-CONECT_emo1}}
    \end{subfigure} 
        %
    \begin{subfigure}{0.8\linewidth}
        \includegraphics[width=\linewidth]{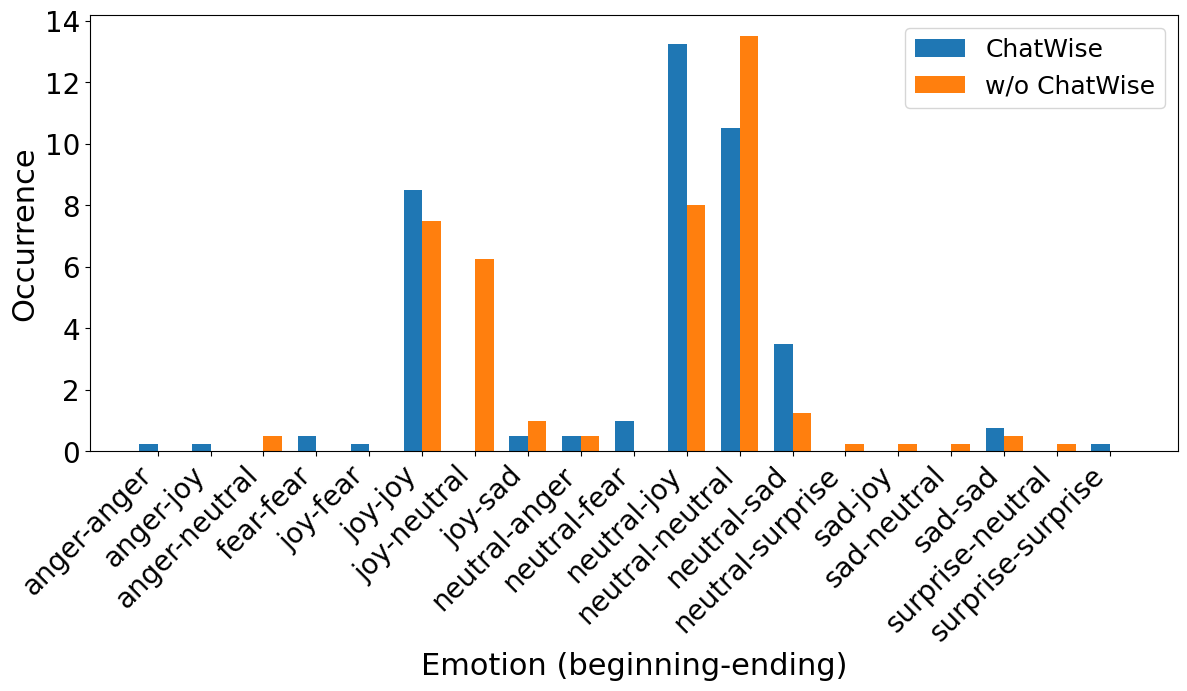}
        \captionsetup{justification=centering}
        \vspace{-0.15in} 
        \caption{\textbf{\textit{Digital twin}} emotion transitions summarized over \J{80 dialgoues}.
    \label{fig:emo_change}}
    \end{subfigure}  
    \caption{User emotion shifts in conversations between digital-twin and \sysName\ (Figure \ref{fig:emo_change}) align well with real-user emotion shifts in the clinical study (Figure \ref{fig:I-CONECT_emo1}). More \textit{neutural-to-joy} transitions were observed given \sysName\ compared to the baseline.}
\vspace{-0.2in}
\end{figure}

\paragraph{User Emotions over Multi-Turn Conversations:} 
We then analyzed user emotion changes from the beginning to the end of dialogues and calculated averaged occurrence across digital twins (Figure \ref{fig:emo_change}).
Over 48\%  users experienced emotional shifts post-dialogue, with significantly more positive changes after engaging with \sysName.
This implies that while transient emotions are difficult to track, strategic conversational support may improve user emotions over multi-turn conversations.

\paragraph{Predominant Strategies:} 
We identified the top 10 most frequent strategies across digital twins \Zhengbang{(Figure \ref{fig:insight_DA})} and found that  {predominant interaction strategies remained consistent across user types} (See Appendix \ref{sec:personalized}).
Particularly, \textit{Open Question}, \textit{Statement-non-opinion}, and \textit{Acknowledgment} strategies dominate \sysName\ driven conversations.

\section{Conclusion and Future Work} \vspace{-0.05in}
We present a dual-level framework for AI chatbots that supports multi-turn conversations for older adults by integrating clinical insights into LLM reasoning. 
Extensive evaluations based on both real clinical data and generated conversations showed that our method aligns well with the behavior of professional human caregivers while robustly enhancing the cognitive status of older adults through simulation studies.
Our future work includes training-based methods for further optimizing multi-turn interaction experience and investigating the acceptability and feasibility of our design through real user studies.

\section{Ethical Considerations}
This study involved simulated dialogues using digital twins trained on \textit{de-identified} conversation data from a clinical trial involving older adults. 
Offline analysis of this dataset was conducted to evaluate the similarity between chatbot and caregiver strategies. 
This project did not use any personally identifiable information, and all data processing followed institutional privacy and research ethics guidelines. 
Potential risks are controllable, which include privacy concerns of emotion detection and misinterpretation of conversation due to AI-based evaluation. 
To mitigate this, the emotional detection module in the system is made optional and can be omitted as configured.
\J{Win rate calculations are based on a sanity check where we omitted pair samples that receive inconsistent preference labels by LLM-as-the-judge after swapping their positions.}
We have manually cross-validated a sufficient subset of AI-generated analyses to reduce bias. A real-user study is planned for future validation and will undergo full Institutional Review Board (IRB) approval.

\clearpage
\section*{Limitations} 
All digital twins were provided as fine-tuned GPT-3.5 APIs, and their high economic cost, at the time when this project was going on, constrained the volume of dialogues we generated. To mitigate this, future work will explore accessing digital twins using LLaMA 3.1 to reduce sampling costs while maintaining conversational quality. 
\J{
LLM evaluations have been discussed to show position bias. Regarding win rate calculation, after we took mitigation methods and filtered out pairs with inconsistent preference labels, the remaining samples contributed to 81\% of the data before filtering.
Future mitigation methods will include learning a reward function for more fine-grained and robust evaluation.  }
Furthermore, our study relies either on offline static data or simulated dialogues rather than dynamic user interactions. A real user study is needed to validate the system’s acceptability and feasibility in real-world settings, which we plan to incorporate into our future research.
\bibliography{acl_latex}

\clearpage
\newpage
\appendix
\section{Appendix}
\label{sec:appendix}
\subsection{Prompt design}
\label{sec:prompt}
The prompt used in the experiments and structured output design are available at \href{https://anonymous.4open.science/r/ChatWise-8F53}{https://anonymous.4open.science/r/ChatWise-8F53}, including:
\begin{itemize}
    \vspace{-0.1in}
    \item Strategy provider system prompt.
    \vspace{-0.1in}
    \item Moderator initial system prompt.
    \vspace{-0.1in}
    \item Moderator system prompt with strategies.
    \vspace{-0.1in}
    \item Strategy provider system prompt for ablation study.
    \vspace{-0.1in}
    \item Moderator system prompt with strategies for ablation study.
    \vspace{-0.1in}
    \item Structured output class for OpenAI models as strategy provider.
    \vspace{-0.1in}
    \item Structured output class for OpenAI models as strategy provider in ablation study.
    \vspace{-0.1in}
    \item System prompt for GPT-4o to extract the strategy given by Llama3.1.
\end{itemize}

\subsection{Offline Data Preparation}
We extracted dialogue content from the I-CONECT video conversations and processed it into OpenAI-compatible dialogue history format. Dialogues with fewer than 40 turns were excluded. We then subsampled two subsets for evaluation purposes. To assess the overall strategy alignment of ChatWise, we randomly sampled 150 dialogues. To evaluate the robustness of ChatWise across different participants and time periods, we randomly selected 7 participants who completed the full study and sampled one dialogue from each of their sessions during weeks 1, 9, 17, 25, 33, and 41. All user identifiers and personally identifiable information have been anonymized.

The OpenAI-compatible dialogue history format example:[\{"role": "system", "content": system\_prompt\}, \{"role": "user", "content": user\_content1\}, \{"role": "assistant", "content":assistant\_content1\}, \{"role": "user", "content": user\_content2\}, \{"role": "assistant", "content": assistant\_content2\}, ...].

\subsection{Data Generation Configuration}

\begin{figure*}[!htbp]
    \centering
    \includegraphics[width=0.75\linewidth]{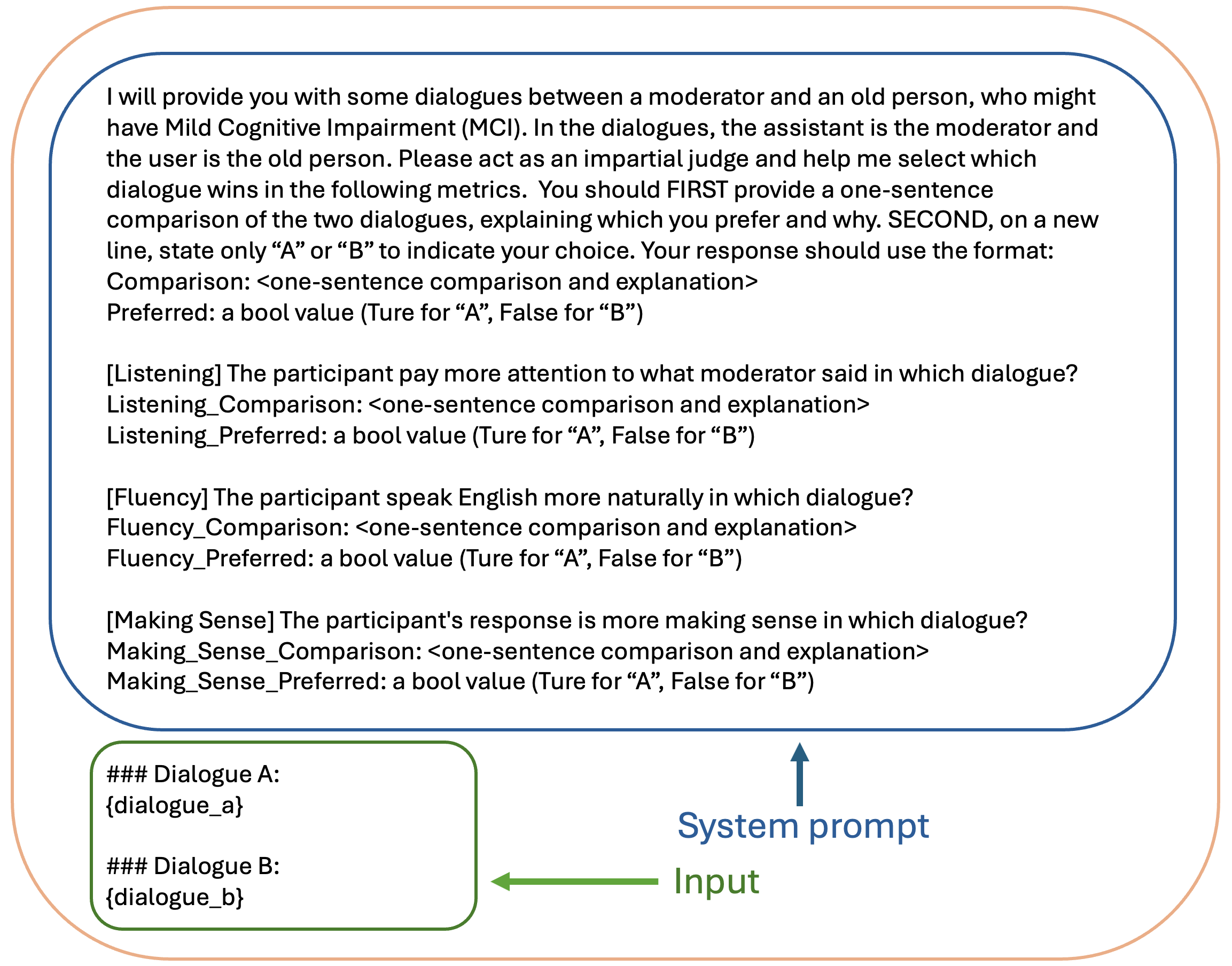}\vspace{-0.1in}
    \caption{Prompt for GPT-4o as a Judge.}
    \label{fig:winrate_prompt}
\end{figure*}

By default, we used o3-mini as strategy provider. GPT-4o serves as an utterance generator in all settings. W/o \sysName\ denotes the baseline, which is using GPT-4o as the utterance generator only, without a strategy provider. We tested different LLM backbones as strategy providers in \sysName. Considering the Llama3.1-405B does not support structured output, we applied GPT-4o as the strategy extractor to structure the strategy output of Llama3.1-405B. We employed GPT-4o as the judge to select the preferred one when computing the Win Rate. The prompt for selecting the preferred response is listed in Figure \ref{fig:winrate_prompt}.
The following are the configurations of each model:

\label{sec:data_gen}
\noindent\textbf{Utterance generator (GPT-4o):}\\
n=1\\
max\_tokens=1024\\
top\_p=1\\
temperature=1\\
\noindent\textbf{GPT-4o, o3-mini as strategy provider:}\\
n=1\\
max\_tokens=1024\\
top\_p=1\\
temperature=1\\
response\_format=Strategy\\
\noindent\textbf{Llama3.1-405B as strategy provider:}\\
top\_p: 0.9\\
max\_tokens: 1024\\
temperature: 0.6\\
presence\_penalty: 0\\
frequency\_penalty: 0\\
\noindent\textbf{GPT-4o as strategy extractor:}\\
n=1\\
max\_tokens=1024\\
top\_p=1\\
temperature=1\\
\noindent\textbf{GPT-4o as judge:}\\
n=1\\
max\_tokens=1024\\
top\_p=1\\
temperature=1\\

\subsection{Win rate for w/o \sysName}
\label{sec:WR_baseline}
The Win rate for w/o \sysName\ is defined as:
\begin{equation*}
\begin{split}
1 - \text{average}(&\text{WR}_{GPT-4o} \\
&+ \text{WR}_{o3-mini} \\
&+ \text{WR}_{Llama3.1-405B})
\end{split}
\end{equation*}
,where $\text{WR}_{X}$ is \sysName's Win rate against the baseline with $X$ as strategy provider.

\subsection{Dialogue Acts} \label{appendix:DA}
The map of strategy to its corresponding abbreviated tag is listed in Table\ref{table:DA_map}.

There are two kinds of strategies: backward-looking and forward-looking. Backward-looking strategies reflect how the current utterance relates to the previous discourse. Forward-looking strategies reflect the current utterance constrains the future beliefs and actions of the participants and affects the discourse. Table \ref{tab:backward-looking DA} and Table \ref{tab:forward-looking DA} provide definitions and examples of each.

\subsection{Log-normalization}
\label{sec:log-normalization}
The following is the log-normalization function, where y is the normalized result, x is the input variable.
\begin{equation*}
    y=\ln{\left(4x+1\right)}
\end{equation*}

\subsection{Primary Strategies}
\label{sec:primary_strategies}
\begin{figure}[!htbp]
    \centering
    \includegraphics[width=0.8\linewidth]{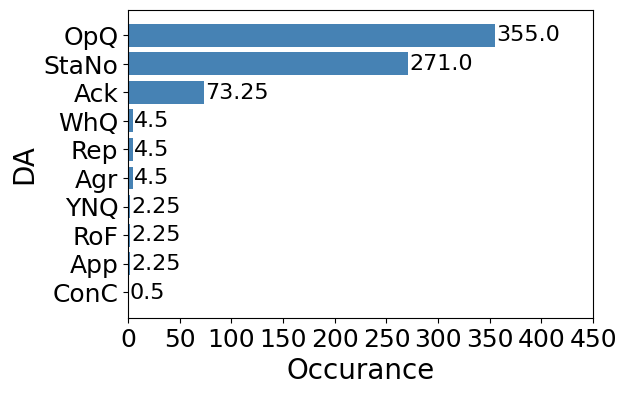}\vspace{-0.1in}
    \caption{Strategy occurrence across digital twins.}
    \label{fig:insight_DA}
\end{figure}

We calculated the average occurrence of each strategy across each digital twin and listed their top 10 most frequently occurring strategies, as shown in Figure \ref{fig:insight_DA}.

\begin{table}[!h]
\vspace{-0.2in}
  \centering
    \begin{tabular}{{p{0.75\linewidth} p{0.25\linewidth} }}
    \hline
    Strategy & Tag\\
    \hline
    Acknowledge (Backchannel) & Ack\\
    Statement-non-opinion & StaNo\\
    Statement-opinion & Sta\\
    Affirmation and Reassurance & Agr\\
    Appreciation & App\\
    Conventional-closing & ConC\\
    Hedge & H\\
    Other & Oth\\
    Quotation & Quo\\
    Action-directive & AcD\\
    Collaborative Completion & CoC\\
    Restatement or Paraphrasing & Rep\\
    Offers Options Commits &Off\\
    Self-talk & Sel\\
    Apology & Apo\\
    Reflection of Feelings & RoF\\
    Yes-No-Question & YNQ\\
    Wh-Question & WhQ\\
    Declarative Yes-No-Question & DYNQ\\
    Open-Question & OpQ\\
    Or-Clause & OrC\\
    Conventional-opening & CoO\\
    Self-disclosure & Sd\\
    Providing Suggestions & PS\\
    Information & I\\
    \hline
    \end{tabular}
  \caption{Strategy to its corresponding tag. The strategies are drawn from the DAs in DAMSL ~\citep{allen1997draft} that are used by telehealth clinical trials ~\citep{YUAN2023100384},  integrated strategies from prior emotional support dataset ~\citep{liu-etal-2021-towards} .}
  \label{table:DA_map}
\vspace{-0.2in}
\end{table}


\begin{table*}[!h]
    \vspace{-0.2in}
  \centering
   \resizebox{0.85\textwidth}{!}{
      \begin{tabular}{{p{0.1\textwidth} p{0.6\textwidth} p{0.3\textwidth}}}
        \hline
        \textbf{Strategy}           & \textbf{Definiton} & \textbf{Example} \\
        \hline
        StaNo
        & A factual statement or descriptive utterance that does not include an opinion. 
        & Me, I'm in the legal department.   \\
        Ack
        & A brief utterance that signals understanding, agreement, or active listening.
        & Uh-huh.\\
        Sta
        & A statement that conveys a personal belief, judgment, or opinion.
        & I think it's great\\
        Agr
        & Affirm the help seeker's strengths, motivation, and capabilities and provide reassurance and encouragement.
        & That's exactly it.\\
        App
        & An expression of gratitude, admiration, or acknowledgment of another’s effort or input. 
        & I can imagine.\\
        ConC
        & A formal or socially standard utterance signaling the end of a conversation. 
        & Well, it's been nice talking to you.\\
        H
        & An expression that introduces uncertainty or qualification to a statement, often to soften its impact. 
        & I don't know if I'm making any sense or not.\\
        Oth
        & Exchange pleasantries and use other support strategies that do not fall into the above categories. 
        & Well give me a break, you know.\\
        Quo
        & A direct or indirect repetition of someone else’s words. 
        & Albert Einstein once said, “Imagination is more important than knowledge.”\\
        AcD
        & A command, request, or suggestion directing someone to take action.
        & Why don't you go first\\
        CoC
        &A continuation or completion of someone else’s utterance in a collaborative manner. 
        & If we want to make it to the top of the mountain before sunset, we should…\\
        Rep
        & A simple, more concise rephrasing of the help-seeker's statements that could help them see their situation more clearly. 
        & It sounds like you’re saying that you’re struggling to stay on top of your work, and it’s leaving you feeling overwhelmed.\\
        Off
        & A statement proposing choices, making a commitment, or offering to do something. 
        & I'll have to check that out\\
        Sel
        & An utterance directed at oneself, often reflecting internal thought processes or problem-solving.
        & What's the word I'm looking for\\
        Apo
        & An expression of regret or asking for forgiveness. 
        & I'm sorry.\\
        RoF
        & Articulate and describe the help-seeker's feelings.
        & It sounds like you’re feeling really frustrated and drained because your efforts don’t seem to be paying off.\\
        \hline
      \end{tabular}
  }
  \caption{Backward-looking strategies, definition, and example. }
  \label{tab:backward-looking DA}
\end{table*}

\begin{table*}[!htbp]
  \centering
   \resizebox{0.85\textwidth}{!}{
      \begin{tabular}{{p{0.1\textwidth} p{0.5\textwidth} p{0.4\textwidth}}}
        \hline
        \textbf{Strategy} & \textbf{Definiton} & \textbf{Example} \\
        \hline
        YNQ
        & A question expecting a binary (yes/no) response. 
        & Do you have to have any special training?\\
        WhQ
        & A question beginning with a wh-word (e.g., what, who, where), seeking specific information. 
        & Well, how old are you?\\
        DYNQ
        & A statement posed as a question, expecting a yes/no answer.
        & So you can afford to get a house?\\
        OpQ
        & A broad question inviting a wide range of responses, often conversational.
        & How about you?\\
        OrC
        & A question offering explicit alternatives, often in the form of “or.”
        & or is it more of a company?\\
        CoO
        & A socially standard utterance used to initiate a conversation.
        & How are you?\\
        Sd
        & Divulge similar experiences that you have had or emotions that you share with the help-seeker to express your empathy.
        & I completely understand how you feel. I remember feeling the same way before my first big presentation at work. I was so anxious, but I found that practicing a few extra times really helped calm my nerves.\\
        PS
        & Provide suggestions about how to change, but be careful to not overstep and tell them what to do.
        & You can keep a note to stop your idea from going. \\
        I
        & Provide useful information to the help-seeker, for example with data, facts, opinions, resources, or by answering questions.
        & Taking silver line from Washington D.C. to Dulles Intel Airport costs about 1 hour.\\
        \hline
      \end{tabular}
  }
  \caption{Forward-looking strategies, definition, and example. }
  \label{tab:forward-looking DA}
\end{table*}
\clearpage
\newpage


\subsection{Personalized Dialogue Analysis}
\label{sec:personalized}
\begin{figure}[!h]
    \centering
    \begin{subfigure}{0.8\linewidth}
        \includegraphics[width=\linewidth]{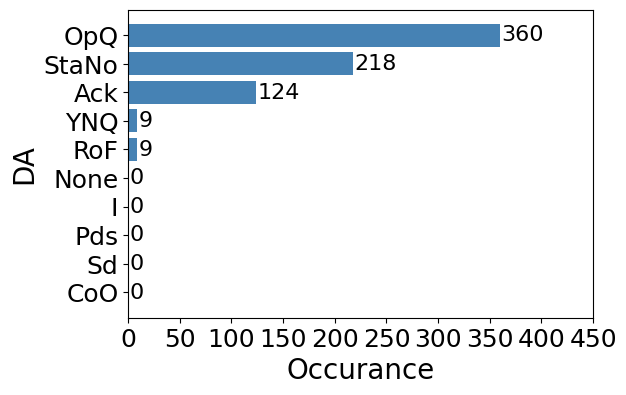}
        \captionsetup{justification=centering}
        \vspace{-0.25in}\caption{Digital Twin 3.}
    \end{subfigure} \hfill
    \begin{subfigure}{0.8\linewidth}
        \includegraphics[width=\linewidth]{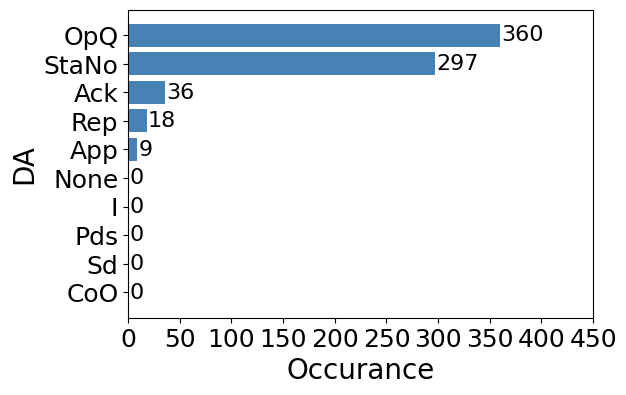}
        \captionsetup{justification=centering}
        \vspace{-0.25in}\caption{Digital Twin 5.}
    \end{subfigure} \hfill
    \begin{subfigure}{0.8\linewidth}
        \includegraphics[width=\linewidth]{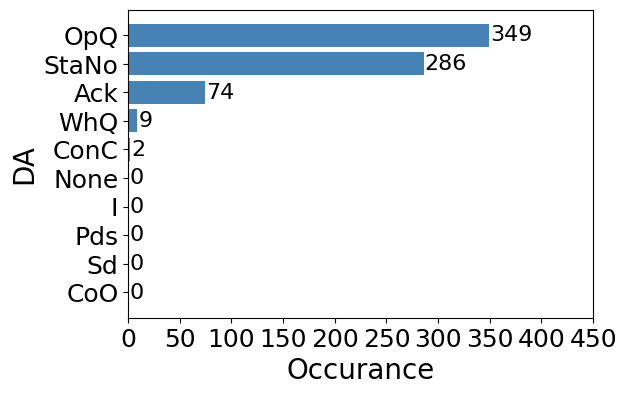}
        \captionsetup{justification=centering}
        \vspace{-0.25in}\caption{Digital Twin 6.}
    \end{subfigure} \hfill
    \begin{subfigure}{0.8\linewidth}
        \includegraphics[width=\linewidth]{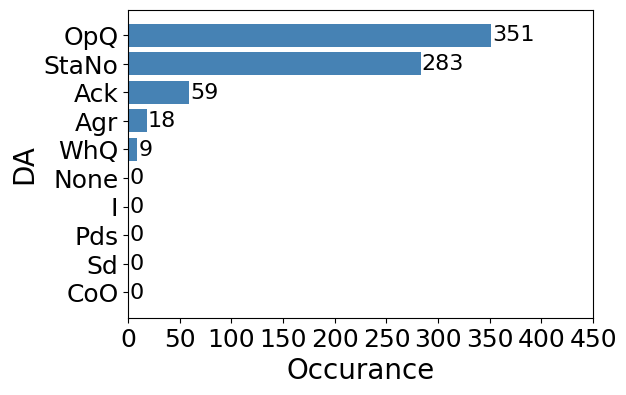}
        \captionsetup{justification=centering}
        \vspace{-0.25in}\caption{Digital Twin 9.}
    \end{subfigure}
    \vspace{-0.1in}
    \caption{Strategy occurrence across digital twins.}
    \label{fig:DA_occ_digital_twin}
    \vspace{-5in}
\end{figure}

\begin{figure}[!t]
    \vspace{-2.5in}
    \centering
    \begin{subfigure}{\linewidth}
        \includegraphics[width=\linewidth]{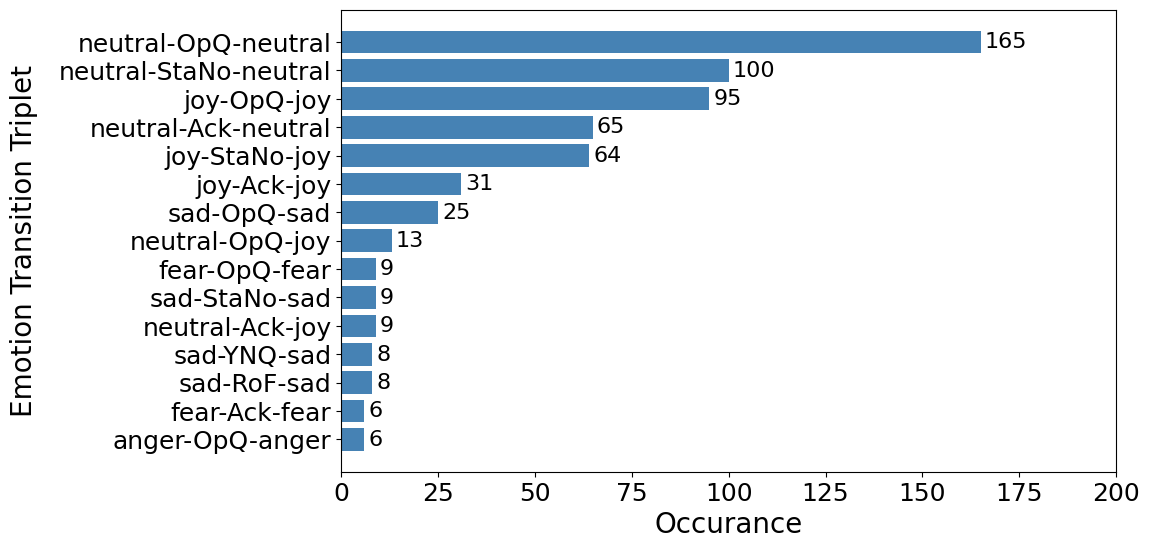}
        \captionsetup{justification=centering}
        \vspace{-0.25in}\caption{Digital Twin 3.}
    \end{subfigure} \hfill
    \begin{subfigure}{\linewidth}
        \includegraphics[width=\linewidth]{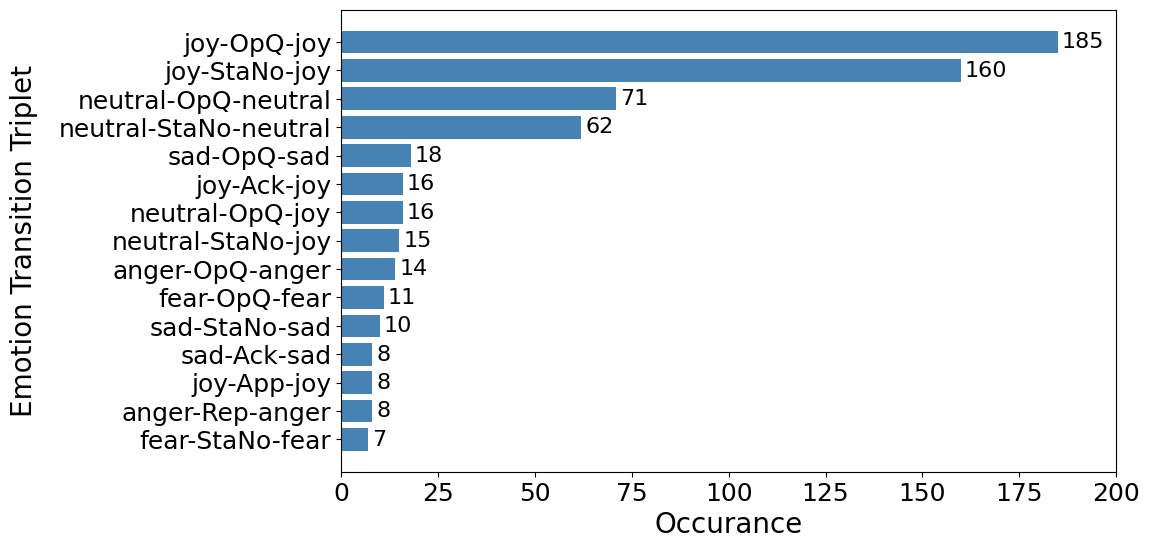}
        \captionsetup{justification=centering}
        \vspace{-0.25in}\caption{Digital Twin 5.}
    \end{subfigure} \hfill
    \begin{subfigure}{\linewidth}
        \includegraphics[width=\linewidth]{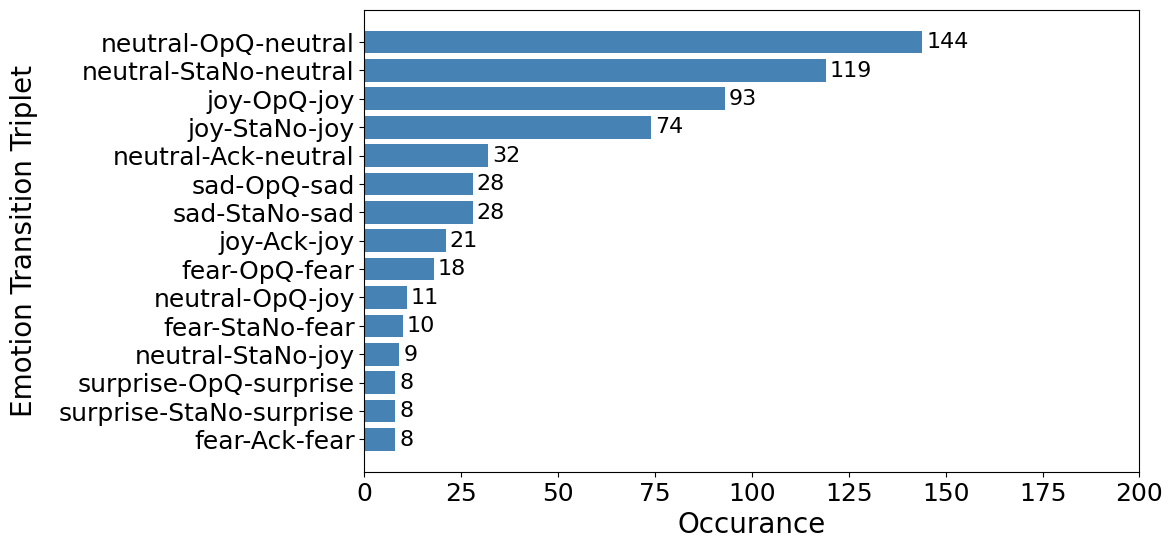}
        \captionsetup{justification=centering}
        \vspace{-0.25in}\caption{Digital Twin 6.}
    \end{subfigure} \hfill
    \begin{subfigure}{\linewidth}
        \includegraphics[width=\linewidth]{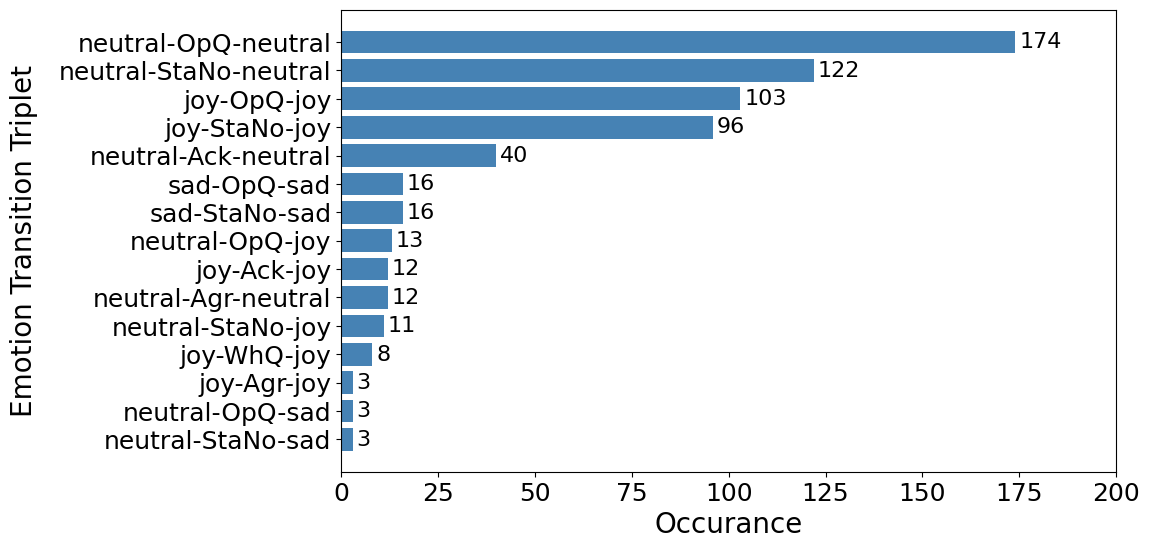}
        \captionsetup{justification=centering}
        \vspace{-0.25in}\caption{Digital Twin 9.}
    \end{subfigure}
    \vspace{-0.25in}
    \caption{Occurrence of each emotion transition triplet across digital twins.}
    \label{fig:DA_emo_digital_twin}
\end{figure}

This section shows the personalized dialogue analysis. The Strategy occurrence across digital twins is shown in Figure \ref{fig:DA_occ_digital_twin}. The Occurrence of each emotion transition triplet across digital twins is shown in Figure \ref{fig:DA_emo_digital_twin}.
%
The \textit{Open Question}, \textit{Statement-non-opinion}, and \textit{Acknowledgment} strategies still dominate \sysName\ driven conversations, suggesting their potential effectiveness in fostering engagement in conversations.
The user's emotion is detected as unchanged in most triplets, indicating the difficulty of altering or measuring the user’s emotional movement within a short turn.

\end{document}